\documentclass[12pt]{article}
\usepackage{times}
\usepackage{geometry}
\usepackage[utf8]{inputenc}
\usepackage{wrapfig}
\usepackage{enumitem,amssymb,graphicx}
\usepackage[numbers]{natbib}
\usepackage{ragged2e}
\newlist{thematic}{itemize}{8}
\setlist[thematic]{label=$\square$}
\usepackage{pifont}
\newcommand{\cmark}{\ding{51}}%
\newcommand{\done}{\rlap{$\square$}{\raisebox{2pt}{\large\hspace{1pt}\cmark}}%
\hspace{-2.5pt}}

\newcommand{\captionfonts}{\footnotesize}

\makeatletter
\long\def\@makecaption#1#2{%
  \vskip\abovecaptionskip
  \sbox\@tempboxa{{\captionfonts #1: #2}}%
  \ifdim \wd\@tempboxa >\hsize
    {\captionfonts #1: #2\par}
  \else
    \hbox to\hsize{\hfil\box\@tempboxa\hfil}%
  \fi
  \vskip\belowcaptionskip}

\renewcommand{\section}{\@startsection%
{section}{1}{0mm}{-0.25\baselineskip}%
{0.05\baselineskip}{\normalfont\normalsize\bfseries}}%

\makeatother

\newcommand{\chandra}{\textit{Chandra}}
\newcommand{\Msun}{\ensuremath{M_\odot}}

\newcommand{\apj}{\textit{ApJ}}

\newcommand{\mnras}{\textit{MNRAS}}

\newcommand{\apjl}{\textit{ApJ}}
\newcommand{\aap}{\textit{A\&A}}

\newcommand{\nat}{\textit{Nature}}

\begin{document}
\raggedright
\huge
Astro2020 Science White Paper \linebreak

The Most Powerful Lenses in the Universe: Quasar Microlensing as a Probe of the Lensing Galaxy \linebreak
\normalsize

\noindent \textbf{Thematic Areas:} \hspace*{60pt} $\square$ Planetary Systems \hspace*{10pt} $\square$ Star and Planet Formation \hspace*{20pt}\linebreak
$\square$ Formation and Evolution of Compact Objects \hspace*{31pt} $\square$ Cosmology and Fundamental Physics \linebreak
  $\square$  Stars and Stellar Evolution \hspace*{1pt} $\square$ Resolved Stellar Populations and their Environments \hspace*{40pt} \linebreak
  $\done$  Galaxy Evolution   \hspace*{45pt} $\square$             Multi-Messenger Astronomy and Astrophysics \hspace*{65pt} \linebreak
  
\textbf{Principal Author:}

Name: David Pooley
 \linebreak						
Institution: Trinity University
 \linebreak
Email: dpooley@trinity.edu
 \linebreak
Phone: (210)\,999-7545
 \linebreak
 
\textbf{Co-authors:} 
Timo Anguita (Universidad Andres Bello), 
Saloni Bhatiani (University of Oklahoma),
George Chartas (College of Charleston),
Matthew Cornachione (United States Naval Academy),
Xinyu Dai (University of Oklahoma),
Carina Fian (Instituto de Astrof\'{i}sica de Canarias; Universidad de la Laguna), 
Evencio Mediavilla (Instituto de Astrof\'{i}sica de Canarias; Universidad de la Laguna), 
Christopher Morgan (United States Naval Academy),
Ver\'{o}nica Motta (Instituto de F\'{i}sica y Astronom\'{i}a, Universidad de Valpara\'{i}so), 
Leonidas~A.~Moustakas (Jet Propulsion Laboratory, California Institute of Technology),
Sampath Mukherjee (University of Groningen),
Matthew J.\ O'Dowd (Lehman College, City University of New York), 
Karina Rojas (Ecole Polytechnique F\'{e}d\'{e}rale de Lausanne; LSSTC Data Science Fellow),
Dominique Sluse (STAR institute, Universit{\'e} of Li{\`e}ge),
Georgios Vernardos (University of Groningen), and
Rachel Webster (University of Melbourne)
\linebreak

\textbf{Abstract  (optional):}
Optical and X-ray observations of strongly gravitationally lensed quasars (especially when four separate images of the quasar are produced) determine not only the amount of matter in the lensing galaxy but also how much is in a smooth component and how much is composed of compact masses (e.g., stars, stellar remnants, primordial black holes, CDM sub-halos, and planets).  Future optical surveys will discover hundreds to thousands of quadruply lensed quasars, and sensitive X-ray observations will unambiguously determine the ratio of smooth to clumpy matter at specific locations in the lensing galaxies and calibrate the stellar mass fundamental plane, providing a determination of the stellar $M/L$.  A modest observing program with a sensitive, sub-arcsecond X-ray imager, combined with the planned optical observations, can make those determinations for a large number (hundreds) of the lensing galaxies, which will span a redshift range of $\sim$$0.25<z<1.5$

\pagebreak
\justifying
\section{Introduction to Microlensing}
Nature has given us a tool more powerful than any telescope we could build in the next several decades.  This tool is the combination of strong gravitational lensing by a galaxy and further lensing by the individual masses inside the galaxy.  

When a massive galaxy happens to lie between us and a distant quasar, it gravitationally lenses the quasar's light, producing multiple observed images (two or four) of the quasar, and introducing a magnification to each one. The locations of the images allow us to determine the amount of mass responsible for the lensing, and the multiplicity of the images can be understood in terms of Fermat's principle, namely, that light will take paths that correspond to stationary points of the travel time; we observe the images that correspond to minima and saddle points of the travel time (two of each in the case of quadruply lensed quasars). 

\begin{figure}[h!]
    \centering
    \includegraphics[width=0.9\textwidth]{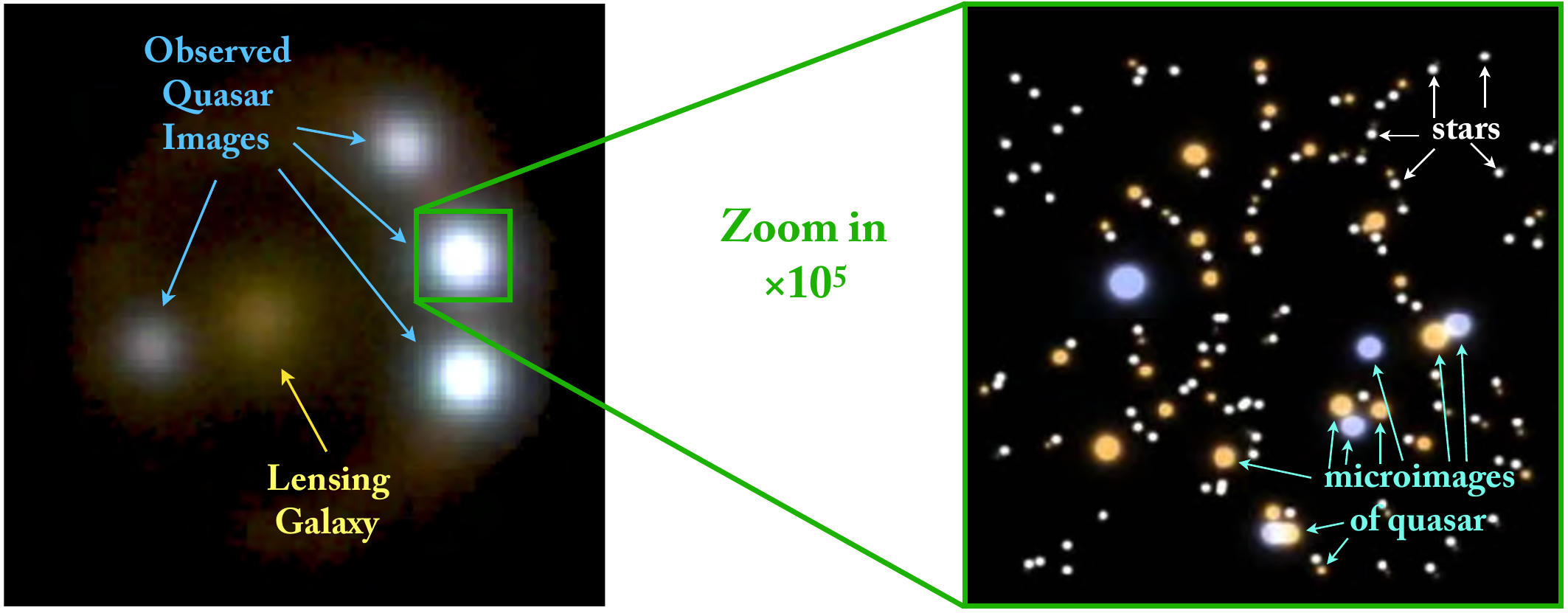}
    \caption{{\it Left:} Magellan image (6$''\!\times\! 6''$) of RX\,J1131$-$1231. {\it Right:} Simulated 60 $\mu$as $\times$ 60 $\mu$as region of the lensing galaxy where the macroimage forms.  The background quasar is located in the center of the image, and this is where a single image would be formed if the mass distribution were completely smooth (i.e., no stars). The microlensing stars are shown in white and the quasar microimages formed by the stars in red (saddle-points) and blue (minima).  Arrows point to a few examples of each.  The single macroimage in the green box on the left is the sum of all the individual, unresolved microimages shown on the right.}
    \label{fig:lensing}
\end{figure}

Each of these observed images of the quasar (henceforth, a ``macroimage'') is in fact the sum total of multiple {\it microimages} of the quasar formed by the stars in the lensing galaxy (Fig.~\ref{fig:lensing}).  We call these stars ``microlenses,'' and they have characteristic Einstein radii of microarcseconds ($\mu$as).  They are the most powerful zoom lenses in the universe.

\begin{figure}[t]
    \centering
    \includegraphics[width=\textwidth]{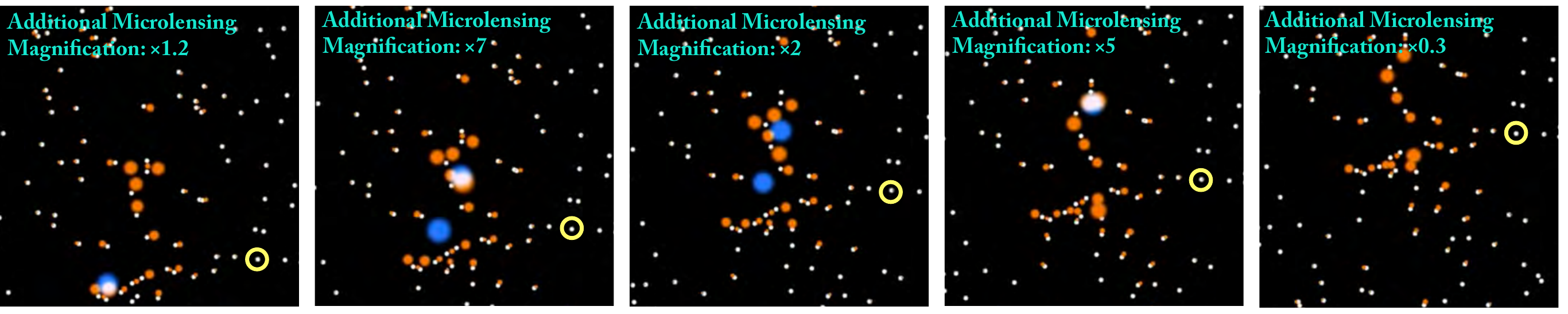}
    \caption{Five views of the microimages of a quasar.  In each, the background quasar is located in the center of the image, and the microimages formed by the stars are shown in red (saddle-points) and blue (minima).  The star field is shifted slightly from one image to the next, and one star is marked with a yellow circle to help guide the eye.  Over an extremely small change in relative position (10s of $\mu$as from the leftmost image to the rightmost), the macroimage (which is the sum of all microimages) would experience changes in brightness by over an order of magnitude.}
    \label{fig:starfields}
\end{figure}

Both the number of microimages and their brightnesses are a sensitive function of the position of the background quasar relative to the small patch of microlenses that forms the macroimage.  
As the quasar and lensing galaxy move relative to each other, the observed brightness of a single macroimage can vary greatly on a timescale of months or years (Fig.~\ref{fig:starfields}).  This variability has nothing to do with any intrinsic variability of the quasar\footnote{The possibility that intrinsic variability combined with the time delay between different macroimages could masquerade as microlensing variability has been shown to be negligible \citep[e.g.,][]{2007ApJ...661...19P}.  However, studying systems with known time delays does give a cleaner microlensing result and also allows one to break the mass-sheet degeneracy \citep{1985ApJ...289L...1F, 2013A&A...559A..37S}.  Time delays themselves are of great interest in determining cosmological parameters \citep[e.g.,][]{2015ApJ...800...11L, 2016A&A...585A..88B}.}.  It is due to the  motion of the quasar relative to the network of microlenses, and it is one of the key observational signatures of microlensing.

\begin{figure}[t!]
    \centering
    \includegraphics[width=0.9\textwidth]{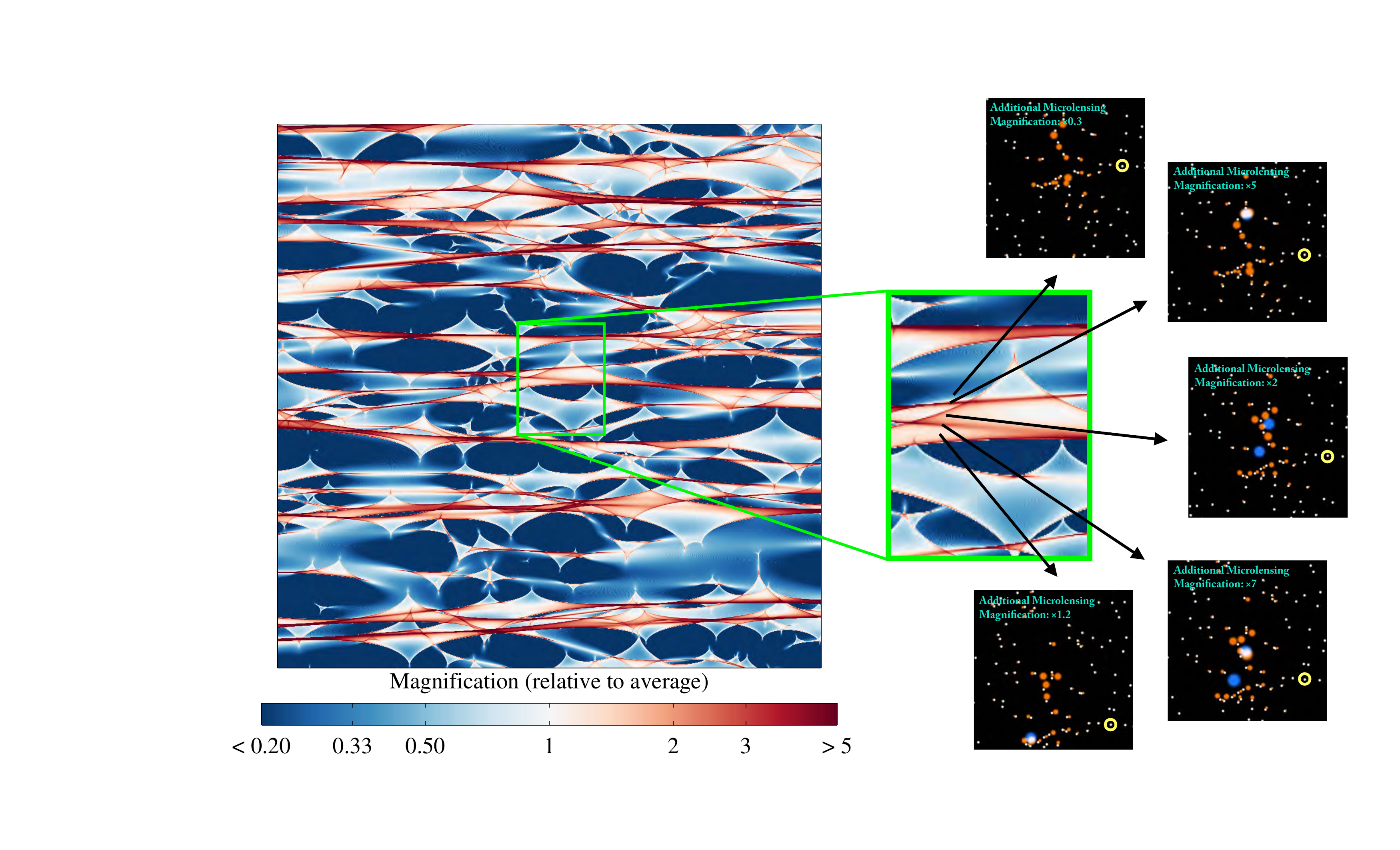}
    \caption{One of the main tools of microlensing analysis is the magnification map, shown on the left, which represents the effects of the entire network of microlenses and shows their perturbative effect, i.e., the additional magnifcation that the microlenses produce on top of the macro-magnification of the lensing galaxy. Each pixel corresponds to a location in the simulated patch of the lensing galaxy and represents the sum total of the microlensing effects (further magnification or even a de-magnification of the quasar) if the quasar were behind that location.  The enlarged area shows the correspondence between five pixels in the magnification map and the five images shown in Fig.~\ref{fig:starfields}.}
    \label{fig:magmaps}
\end{figure}

We cannot resolve the individual microimages or the microlenses themselves. We use realizations of the microlens field to understand the microlensing variability that we observe.  For each location in a small patch of the lensing galaxy, we can determine how magnified the macroimage would be if the background quasar were located behind that position, given the ratio of smooth to clumpy matter.  Each position of the background quasar then corresponds to a particular magnification of the observed macroimage, and the map of this is called a microlensing magnification map (Fig.~\ref{fig:magmaps}).  The most prominent features of these maps are the sharp lines of intense magnification, called ``caustics,'' that correspond to the creation or destruction of a pair of microimages.

Microlensing allows us to probe both the source (the quasar) and the lens (the intervening galaxy) in ways that no other method can accomplish:
\vspace{-1.25ex}
\begin{itemize}
\setlength{\itemsep}{-0.5ex}
    \item As the knife-edge caustic sweeps over the quasar, different emitting regions are lensed.  Multi-wavelength observations of a caustic-crossing event can provide sub-microarcsecond resolution of the structure of quasars at typical redshifts of $z\approx2$, and a companion white paper addresses this.
    \item This white paper focuses on how the frequency of occurrence of microlensing-induced variations depends on the granularity of matter in the lensing galaxy.  One can thus determine the amount of mass in compact form (faint stars, black holes, brown dwarfs, planets, etc.) at specific locations (typically around several kpc from the center) in the lensing galaxy.  
\end{itemize}

\vspace{-.5ex}
Measuring the various {\it invisible} mass constituents of a galaxy is both fundamentally important in understanding galaxies and impossible to do by any other means than microlensing.   One can find papers with hundreds of citations that give stellar mass-to-light ratios for elliptical galaxies, but buried within these papers the authors invariably caution that uncertainties in the faint end of the stellar mass function, where the stars are {\it invisible}, render their results uncertain by a factor of two.  This was summarized as follows:

\vspace{-1.5ex}
\begin{quote}{{\it Nobody ever measures the stellar mass. That is not a measurable thing; it’s an inferred quantity. You measure light, OK? You can measure light in many bands, but you infer stellar mass. Everybody seems to agree on certain assumptions that are completely unproven.} --- Carlos Frenk, 2017 May 15 \footnote{{http://online.kitp.ucsb.edu/online/galhalo-c17/panel1/rm/jwvideo.html} (44:48)}}
\end{quote}

\vspace{-1.5ex}
\noindent Quasar microlensing is the {\it only} way to determine the stellar $M/L$  beyond the solar neighborhood.

\section{Source Size: the Necessity of Sub-arcsecond X-ray Imaging to Determine Stellar Masses}
The ability of any lens to significantly magnify or demagnify a source of light is related to the size of the lens (the Einstein radius for gravitational lenses) compared to the size of the source of light.  The emitting region of a quasar is a function of wavelength, with X-rays coming from a compact region near the black hole, and optical light coming from farther out (in the disk or possibly beyond).  Because the optical comes from a region comparable in size to the microlenses' Einstein radii, the microlensing effects are diminished.  The X-ray-emitting region is essentially a point source and therefore gives a clean microlensing signal, unlike the optical which gives a convolution of microlensing and the finite size of the optical emitting region of the quasar.  {\it High spatial resolution X-ray observations are crucial to using the microlensing of quasars as a tool.}

\begin{figure}[t!]
    \centering
    \includegraphics[width=0.8\textwidth]{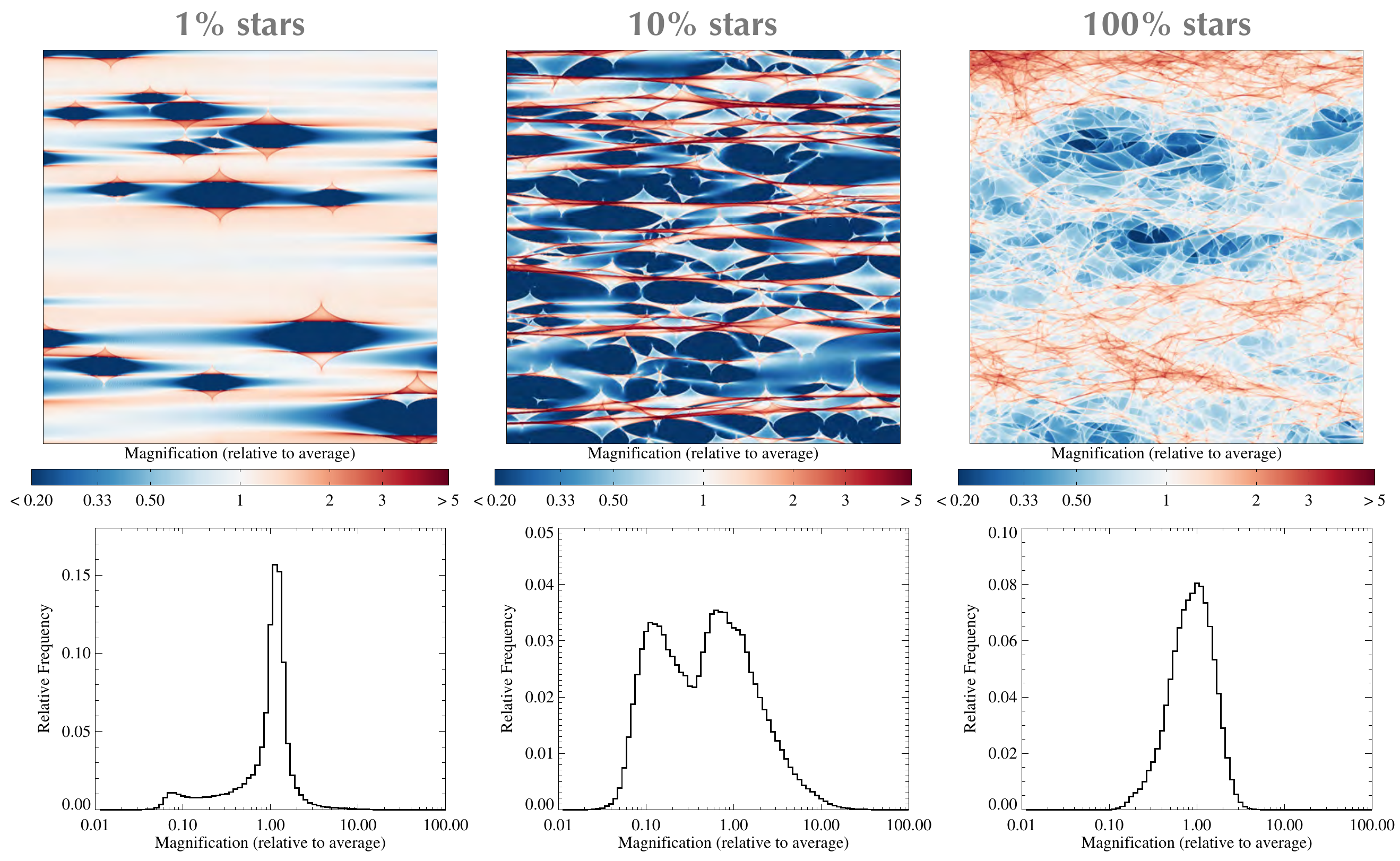}
    \caption{Microlensing magnfication maps for a macro saddle point image are shown for three representative stellar mass fractions.  Below each map are shown the normalized histograms of the map's pixel values, which are probability distributions for observed microlensing effects.  Notice that the probability of a strong demagnification (factor of $\sim$10) is small for both low and high stellar fractions but is appreciable for the intermediate stellar fractions.  In principle, a histogram of {\it observed} macro-image brightness measurements could be compared to these simulated histograms to infer the stellar fraction at the locations of each of the four macroimages formed by the lensing galaxy.  In practice, obtaining statistically independent measurements for such an analysis requires a time baseline of decades.  A future X-ray mission with high sensitivity and sub-arcsecond spatial resolution would leverage the legacy of \chandra's observations of quads to allow for just such a determination. }
    \label{fig:lensingprobs}
\end{figure}

\section{Determining the Smooth/Clumpy Matter Ratio}
\citet{2002ApJ...580..685S} explored the microlensing effects of different fractional contributions of stars and dark matter to the total surface density, and they found that the probability of a strong demagnification of a saddle point image, which is often seen in the observations, was relatively low for stellar fractions of 2\% and 100\% but became appreciable for stellar fractions of 5\%--25\% (e.g., see Fig~\ref{fig:lensingprobs}). Then, using an ensemble of 11 lensing galaxies, Schechter \& Wambsganss determined the most likely stellar fraction at the typical impact parameter of image formation \citep{2004IAUS..220..103S}. They noted, however, that their analysis produced inconsistent results unless they assumed that the optical continuum emitting regions had an extended component. 

The X-rays do not suffer such complications and offer a much more promising avenue, both for individual systems \citep[e.g.,][]{2008ApJ...689..755M} and an ensemble of systems.  The clean signal of microlensing in X-rays was used for the ensemble of known quads to determine a most likely local stellar mass fraction of 7\% at a mean distance of 6.6 kpc from the center of a typical lensing galaxy \citep{2012ApJ...744..111P}.  

\section{The Coming Deluge of Quads: Mass Determinations as a Function of Redshift}
The Large Scale Synoptic Telescope (LSST) is expected to discover thousands of quadruply lensed quasars.  The full power of these discoveries will be unlocked with high spatial resolution X-ray observations.  A facility with similar spatial resolution to \chandra\ and $\sim$50 times the effective area will able to study several hundred of these new discoveries with a modest observing program (several hundred ksec of exposure time).\footnote{Typical current \chandra\ observations of quads have exposures around 10--50 ksec, but the LSST-discovered quads will likely be somewhat fainter than the currently known quads.}  Such a facility could carry out two types of studies:

\vspace{-1.5ex}
\begin{itemize}
\setlength{\itemsep}{-0.75ex}
    \item The legacy of \chandra\ observations would be leveraged by having X-ray microlensing observations separated by decades, allowing for determinations of the smooth/stellar mass components in the individual lensing galaxies currently known.
    \item The ensemble analysis described above could be done on the newly discovered lensing galaxies in redshift bins spanning $\sim\!0.25<z<1.5$, which is where the bulk of the LSST discovered lensing galaxies will reside \citep{2010MNRAS.405.2579O}.  One could then study the evolution of the smooth/stellar mass components and the $M/L$ ratio.
\end{itemize}

\vspace{-1.5ex}

\begin{figure}
    \centering
    \includegraphics[width=\textwidth]{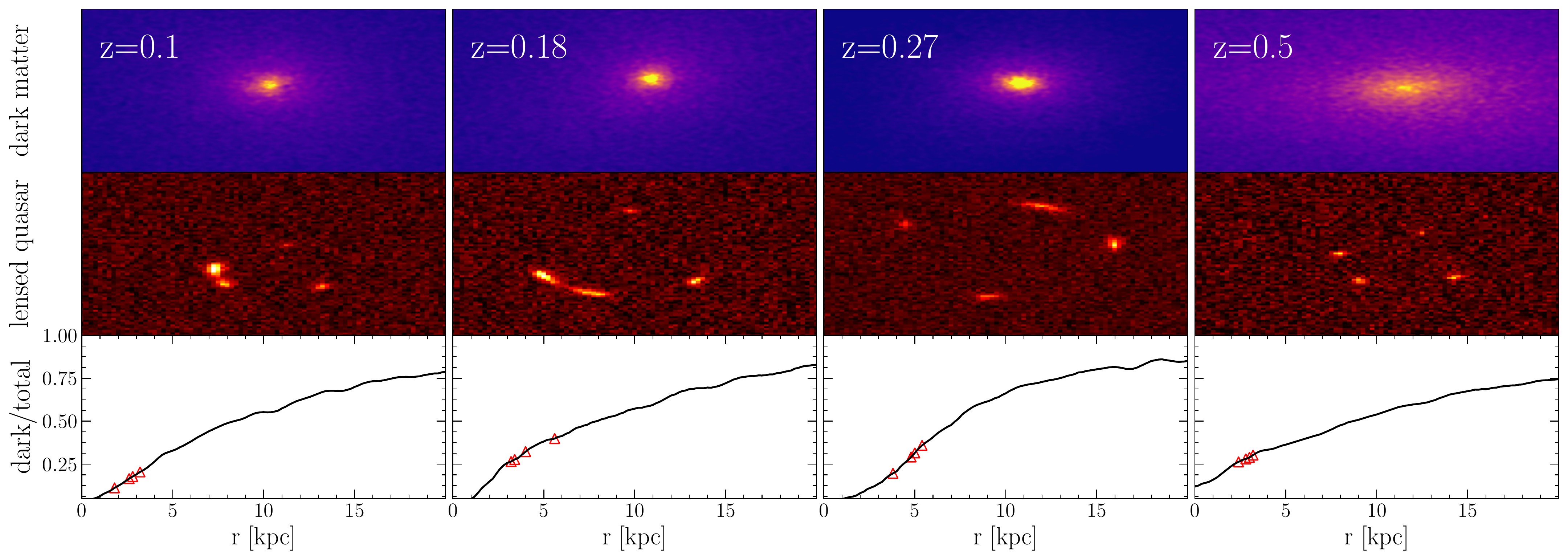}
    \caption{Four fiducial galaxies have been selected from the EAGLE simulations with various galaxy formation scenarios \citep{2015MNRAS.446..521S}. {\it Top:} dark matter profile of the lensing galaxies. {\it Middle:} corresponding quadruply imaged quasar lens configurations (the lens light has been subtracted). {\it Bottom:} the average dark over total matter fraction of the lens galaxies as a function of radius (calculated within elliptical contours aligned with the galaxy mass distribution). Microlensing is the \textit{only} method that can measure the dark over baryonic matter fraction at the locations of the multiple quasar images, indicated by the triangles, within the most interesting sub-galactic scales ($<30$ kpc).}
    \label{fig:dmfrac}
\end{figure}

\section{Stellar \boldmath{$M/L$} via the Fundamental Plane}
The overall mass density of a lensing galaxy is known from the macro-lensing.  X-ray observations determine the level of microlensing, from which one determines the amount of mass in individual stars, including stellar remnants, brown dwarfs, and red dwarfs too faint to produce photometric or spectroscopic signatures.  To determine $M/L$, one needs only to measure the amount of light, but this is problematic where the quasar images form.  One solution is to use the stellar mass fundamental plane \citep{2009MNRAS.396.1171H} and determine a calibration factor $\mathcal{F}$ by which one multiplies the plane (constructed using a Salpeter initial mass function with a low-mass cutoff of 0.1 \Msun) to obtain the best agreement with the observed fluxes. The median likelihood value for the normalization factor $\cal F$ by which the Salpeter stellar masses must be multiplied is 1.23, with a one sigma confidence range, dominated by small number statistics, of $0.77 <{\cal F}< 2.10$ \citep{2014ApJ...793...96S}.

\section{Constraints on Massive Compact Halo Objects (MaCHOs)}

\begin{wrapfigure}{r}{0.42\textwidth}
    \centering
    \vspace{-15pt}
    \includegraphics[width=0.4\textwidth]{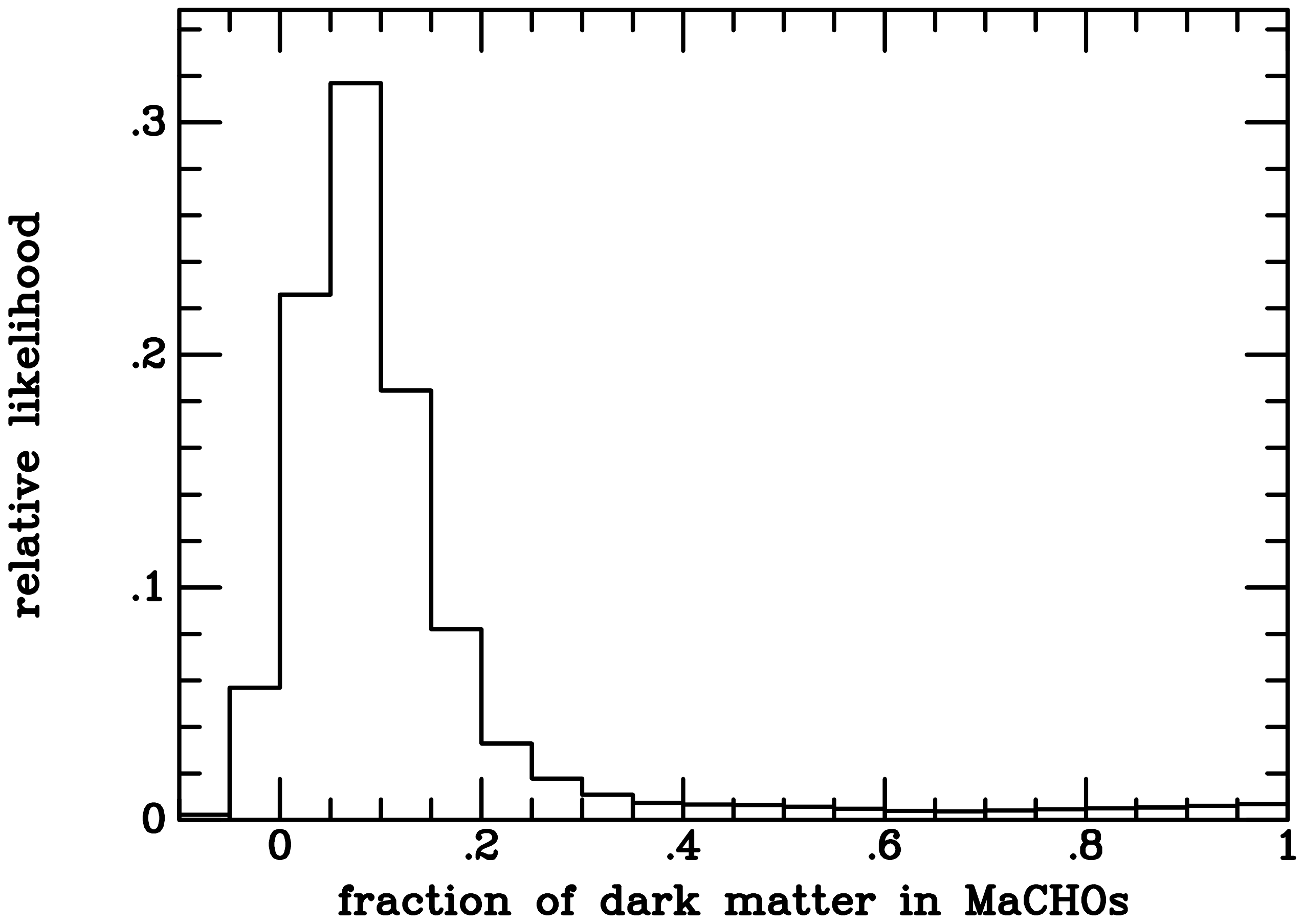}
    \caption{from \citet{2018ASPC..514...79S}: ``Likelihoods for a range of fractional contributions of MaCHOs to the dark matter surface density in ten lensed quasars. Note the finite likelihood for a negative fraction, which would result if a Salpeter IMF overestimates the surface mass density.'' }
    \label{fig:massconstraint}
    \vspace{-10pt}
\end{wrapfigure}

Most studies of microlensing have emphasized the determination of the fraction of mass in compact objects by considering a limited range of masses, preferably close to typical stellar values \citep[e.g.,][]{2004IAUS..220..103S, 2009ApJ...706.1451M, 2012ApJ...744..111P, 2014ApJ...793...96S, 2015ApJ...799..149J, 2015ApJ...806..251J}.  However, \citet{2017ApJ...836L..18M} and \citet{2018ASPC..514...79S} have used the available observations to constrain the fraction of dark matter in compact objects, considering as wide a mass range as possible, and find the fraction of the dark halo in MaCHOs (including $\sim$20\Msun\ primordial black holes) is $\lesssim$10\% (Fig.~\ref{fig:massconstraint}).

X-ray microlensing can also place constraints on the low-mass range of MaCHOs through a study of the frequency of occurrence of Fe-line shifts in the observed quasar spectra \citep{2018ApJ...853L..27D}, which happen too often if only stellar-mass objects are considered and seem to require the presence of $\sim$2000 Moon- to Jupiter-sized objects per star, making up about $10^{-4}$ of the dark matter.  The leading candidates are free-floating planets (ejected, scattered, or stripped due to various process) or primordial black holes.   Therefore, the planet-scale astronomical dark matter can either serve as a probe of star/planet formation and scatter process or fundamental physics in the very early universe in the inflation era.


\end{document}